\documentclass[aps,pra,twocolumn,groupedaddress,showpacs]{revtex4}
\usepackage{graphicx}

\newcommand{\bea}{\begin{eqnarray}}

\newcommand{\eea}{\end{eqnarray}}

\newcommand{\beq}{\begin{equation}}

\newcommand{\eeq}{\end{equation}}

\begin{document}

\def \tr{{\mbox{tr~}}}
\def \ra{{\rightarrow}}
\def \ua{{\uparrow}}
\def \da{{\downarrow}}
\def \be{\begin{equation}}
\def \ee{\end{equation}}
\def \ba{\begin{array}}
\def \ea{\end{array}}
\def \bea{\begin{eqnarray}}
\def \eea{\end{eqnarray}}
\def \nn{\nonumber}
\def \l{\left}
\def \r{\right}
\def \half{{1\over 2}}
\def \etal{{\it {et al}}}
\def \cH{{\cal{H}}}
\def \cM{{\cal{M}}}
\def \cN{{\cal{N}}}
\def \cQ{{\cal Q}}
\def \cI{{\cal I}}
\def \cV{{\cal V}}
\def \cG{{\cal G}}
\def \cF{{\cal F}}
\def \cZ{{\cal Z}}
\def \bS{{\bf S}}
\def \bI{{\bf I}}
\def \bL{{\bf L}}
\def \bG{{\bf G}}
\def \bQ{{\bf Q}}
\def \bK{{\bf K}}
\def \bR{{\bf R}}
\def \br{{\bf r}}
\def \bu{{\bf u}}
\def \bq{{\bf q}}
\def \bk{{\bf k}}
\def \bp{{\bf p}}
\def \bz{{\bf z}}
\def \bx{{\bf x}}
\def \bpsi{{\bar{\psi}}}
\def \tJ{{\tilde{J}}}
\def \W{{\Omega}}
\def \e{{\epsilon}}
\def \lam{{\lambda}}
\def \L{{\Lambda}}
\def \a{{\alpha}}
\def \t{{\theta}}
\def \b{{\beta}}
\def \g{{\gamma}}
\def \D{{\Delta}}
\def \d{{\delta}}
\def \w{{\omega}}
\def \s{{\sigma}}
\def \f{{\varphi}}
\def \x{{\chi}}
\def \e{{\epsilon}}
\def \h{{\eta}}
\def \G{{\Gamma}}
\def \z{{\zeta}}
\def \hatt{{\hat{\t}}}
\def \hn{{\bar{n}}}
\def \vk{{\bf{k}}}
\def \vq{{\bf{q}}}
\def \gk{{\g_{\vk}}}
\def \nd{{^{\vphantom{\dagger}}}}
\def \yd{^\dagger}
\def \av#1{{\langle#1\rangle}}
\def \ket#1{{\,|\,#1\,\rangle\,}}
\def \bra#1{{\,\langle\,#1\,|\,}}
\def \braket#1#2{{\,\langle\,#1\,|\,#2\,\rangle\,}}


\title{Dynamic Kosterlitz-Thouless transition in 2D Bose mixtures of ultra-cold atoms}

\author{L. Mathey$^1$, Kenneth~J.~G\"unter$^2$, Jean~Dalibard,$^2$  and A.~Polkovnikov$^3$}
\affiliation{
$^{1}$Zentrum f\"ur Optische Quantentechnologien and Institut f\"ur Laserphysik, Universit\"at Hamburg, 22761 Hamburg, Germany\\
$^2$Laboratoire Kastler Brossel, CNRS, UPMC, Ecole Normale Sup«erieure, 24 rue Lhomond, F-75005 Paris, France\\
$^3$Department of Physics, Boston University, 590 Commonwealth Ave., Boston, MA 02215}

\date{\today}

\begin{abstract}
  We propose a realistic experiment to demonstrate a dynamic 
  Kosterlitz-Thouless transition in ultra-cold atomic gases in two dimensions. With a numerical implementation of the Truncated  Wigner Approximation we simulate the time evolution of several correlation functions, which can be measured via matter wave interference. We demonstrate that the relaxational dynamics is well-described by a real-time renormalization group approach, and argue that these experiments can   guide the development of a theoretical framework for the understanding of  critical dynamics. 
\end{abstract}


\maketitle

The understanding of non-equilibrium phenomena, in particular dynamic phase transitions, is an open frontier in many-body physics.
While for equilibrium systems a rich variety of methods has been established, non-equilibrium systems are notoriously
difficult to grasp, and a functioning conceptual framework is lacking. Given this state of research, ultra-cold atomic systems can play a crucial role in further understanding many-body dynamics.
In fact, the unprecedented control of well-defined, isolated systems of ultra-cold atoms, has led them to be considered as
'quantum simulators' \cite{feynman1982}: Cold atom systems are manipulated to create paradigmatic model systems of condensed matter physics, such as the Bose-Hubbard model \cite{bakr2010}, spin chains \cite{simon2011}, the unitary Fermi gas \cite{vanhoucke2011}, magnetic systems \cite{struck2011},  the Dirac equation \cite{gerritsma2010} and equilibration in one-dimensional (1D) gases \cite{trotzky2011}. The experimental
measurements are cross-checked with theory, to guide the development of a theoretical framework \cite{qsimtheory}. 
 
In this paper, we propose to apply the concept of quantum simulation for a non-equilibrium setup, in particular for dynamic phase transitions. 
We present an implementation of the
time-dependent renormalization group (RG) description, derived in
 \cite{quenchlong}, and demonstrate that it quantitatively predicts the dynamic evolution across a critical point, by comparing it to a
numerical simulation.
In particular, this proves the universality of the relaxational dynamics in this system, a concept well-established for equilibrium phase transitions, 
 however not developed for dynamics in closed systems. Other RG treatments of non-equilibrium systems were reported in \cite{mitra}.

We consider a system of weakly interacting bosons in two dimensions (2D), for a review see e.g. \cite{2Dreview}, which in equilibrium undergoes a Kosterlitz-Thouless (KT) phase transition as a function of temperature \cite{KT}. This
system has two thermal phases, defined through the long-range behavior
of the two-point correlation function
$G(\br)\equiv\langle\psi^\dagger(0)\psi(\br)\rangle$, where
$\psi(\br)$ is the particle annihilation operator at site $\br$.  At
low temperatures this function decays algebraically, 
$G(\br) \sim |\br|^{-\tau/4}$. The exponent $\tau$ increases
monotonically from zero to $1$ as the temperature is increased from
zero to the critical temperature $T_c$.  Above $T_c$, the functional form of $G(\br)$ changes to exponential
decay, $G(\br)\sim \exp(-|\br|/r_0)$, with some decay length
$r_0$.  This change is due to the deconfinement
of vortex-antivortex pairs, and defines the KT transition.
We trigger this transition dynamically by a quench, as described
below. We find that after an intermediate time the system
develops a metastable state in which the phononic modes have
equilibrated, and $G(\br)$ shows algebraic scaling with an exponent
$\tau$, which can be larger than the critical value $1$. This exponent then
increases slowly in time, until the correlation function changes to exponential scaling, indicating dynamic vortex
deconfinement.  We demonstrate that the real-time evolution of
$\tau(t)$ can be described by a real-time RG approach, by comparing it to
our simulations. This constitutes a conceptually new insight into many-body dynamics.
For the ultimate validation of this theoretical approach we propose a specific experimental setup in this paper.

\begin{figure}
\includegraphics[width=8.4cm]{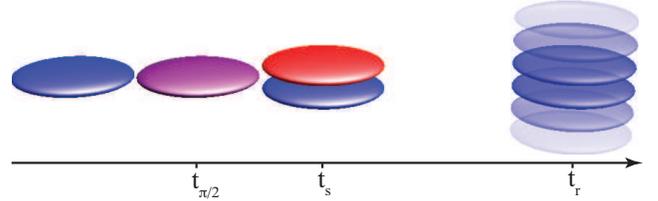}
\caption{\label{sketch} We prepare a 2D atom cloud in state $1$ (blue), and apply a $\pi/2$ pulse at $t_{\pi/2}$. We apply a field gradient at $t_s$, which separates state $1$ and $2$ (red) spatially.  We release the atoms at time $t_r$ and measure their interference properties.}
\end{figure}
\begin{figure*}
\includegraphics[width=16.0cm]{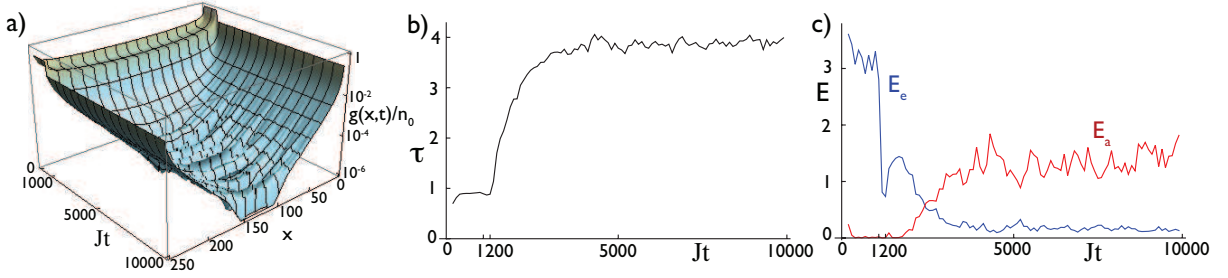}
\caption{\label{spgf}  
 a) $g_1(x,t)/n_0$ on a logarithmic scale, for $U/J=0.25$ and $\tau_i=0.89$ on a $250\times 250$ lattice. 
 The $\pi/2$  pulse is applied at $J t_{\pi/2} = 1000$.  $U_{12}$ is turned off at $J t_s = 1200$. After $t_s$, $g_1(x,t)$ gradually changes
 its functional form, from algebraic to exponential.  b) Fitted algebraic exponent $\tau$ obtained from the data in a).  c) Fitting errors $E_e$ and $E_a$.}
\end{figure*}

In Fig.~\ref{sketch} we sketch the quench and measurement sequence.
 We consider a 2D gas of atoms, such as $^{87}$Rb, at a temperature below $T_c$ with an initial scaling exponent $\tau_i < 1$ in an internal state labelled $1$. At some time $t_{\pi/2}$, a radio-frequency or microwave pulse drives the transition from state $1$ to another state $2$. The duration and intensity of the pulse are adjusted to provide a $\pi/2$ transition, so that the superfluid (SF) is now in the state $(\psi_1(\br) + \psi_2(\br))/\sqrt{2}$, where $\psi_1(\br)$ and $\psi_2(\br)$ are the single-particle  operators of  states $1$ and $2$. If the interaction strengths between $1$ and $2$ are identical (as they approximately are for the hyperfine levels of $^{87}$Rb in its ground state), the rotation between the internal states is a symmetry of the system and the resulting SF state is still a steady state. Then at a later time  $t_s$, we perform a quench of the gas by spatially separating the two species with a magnetic field gradient. This has the effect that  the inter-species interaction is set to zero. For $^{87}$Rb, the two hyperfine states could be  $|F=1, m_F=-1\rangle$ and $|F=2, m_F=-1\rangle$, which have Land\'e factors of opposite sign.  We measure the coherence properties at some release time $t_r$  by (i) switching off the confinement along the initially frozen direction, so that the two gases expand along this direction and overlap; (ii) performing another $\pi/2$ pulse between  $1$ and $2$; (iii) recording the atomic density in one of the two   states, say $1$, which reveals the matter-wave interference between the two planes. We note that 'splitting' of a 1D SF was discussed in  \cite{kitagawa2011}.
 
 We describe the system with the Hamiltonian $H  =  H_1 + H_2 + H_{12}$,   where
\bea
H_a & = & \sum_\br l^D \Big[-\frac{\hbar^2}{2 m} \psi^\dagger_a(\br) \triangle
  \psi_a(\br) -\mu \psi^\dagger_a(\br)  \psi_a(\br)\nonumber\\  
 && + \frac{g_0}{2}  \psi^\dagger_a(\br)  \psi^\dagger_a(\br)  \psi_a(\br) \psi_a(\br)\Big],
\eea
 and 
\bea
H_{12} & = & \sum_\br l^D \Big[g_{12}  \psi^\dagger_1(\br)  \psi^\dagger_2(\br)  \psi_2(\br) \psi_1(\br)\Big],
\eea
where the states are labeled $a= 1,2$, and the dimension $D=2$;   
 $l$ is the discretization length scale of this representation, see Ref.~\cite{mora},   
  $m$ is the atom mass,  $g_0$ ($g_{12}$) is the strength of the
  intra- (inter-) species contact interaction.
 For  atoms confined to 2D motion by a harmonic
 potential in the third direction, these are approximately 
\bea
g_0 & = & \frac{2\sqrt{2\pi} \hbar^2}{m} \frac{a_s}{l_0}\label{g0};
 \, \, \, \, g_{12}  =   \frac{2\sqrt{2\pi} \hbar^2}{m} \frac{a_{12}}{l_0}.
\eea
 $l_0$ is the harmonic oscillator length 
$l_0 = (\hbar/m\omega_0)^{1/2}$  of the confining
 potential $m \omega_0^2 z^2/2$ in $z$ direction; 
 $a_s$ and $a_{12}$ are the s-wave scattering lengths. For 
 hyperfine states of $^{87}$Rb these are around $5$ nm.

 Initially   all atoms are in state $a=1$, and  form a SF of total density $\rho_0$ at a temperature $T<T_c$. We describe this state with a Bogoliubov analysis based on the phase-density representation $\psi(\br)  \approx  e^{i\theta(\br)}\sqrt{\rho_0 + \delta\rho(\br)}$, see Ref.~\cite{mora}.
 As mentioned, the correlation function $G_1(\br)\equiv\langle \psi^\dagger_1(0)\psi_1(\br)\rangle $ decays algebraically, $G_1(\br) \sim |\br|^{-\tau/4}$, with
\bea
\tau^{-1} & = & \pi \hbar^2 \rho_s /(2 mT)\label{tau_general},
\eea
see Refs.~\cite{NP01,NP02}, implicitly defining the SF density $\rho_s$.  Within the Bogoliubov approximation,  $\rho_s=\rho_0$.  However, based on Ref.~\cite{NP02}, for weak interactions and away   from the critical regime, an improved estimate is  
\bea
\tau^{-1} & \approx & \pi \hbar^2 \rho_0/(2 mT) + C_0,\label{tau_approx}
\eea
see  \cite{SFdensity}, with  $C_0 = (\ln(2 g m/\hbar^2) -1)/4$. 
\begin{figure*}
\includegraphics[width=14.7cm]{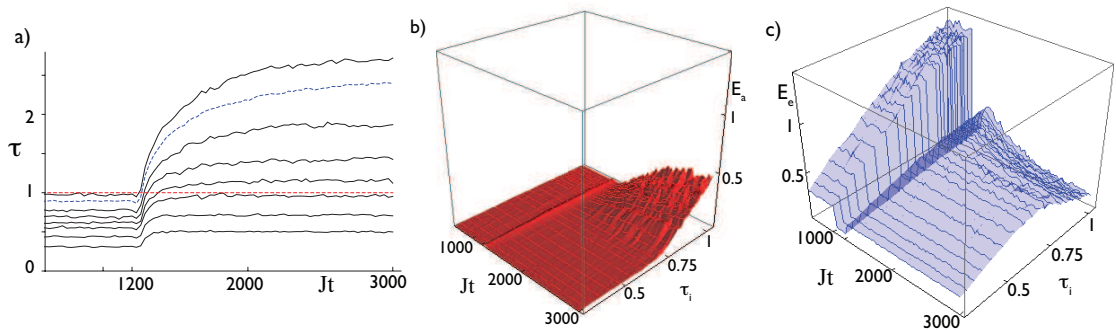}
\caption{\label{tautime}
 a)  Algebraic exponent $\tau (t)$ from fitting $g_1(\bx, t)$ on a $100\times 100$ lattice, for different $\tau_i$. The red, dotted line is the critical value $1$. The blue, dashed line is $\tau(t)$ for $\tau_i=0.89$ (as in Fig.~\ref{spgf}). 
  b) Algebraic and  c) exponential fitting errors $E_{a,e}$. }
\end{figure*}
 We simulate the dynamics with a numerical
 implementation of the Truncated Wigner approximation (TWA), see e.g. Refs.~\cite{blakie, ap_twa}.
  It is convenient to represent the system as 
 a Hubbard model with two species $b_{a,i}$, where 
 $i$ is the lattice site, and $a=1,2$ the species index.
  The single-particle operators of the two representations are related as 
 $b_{a,i}  =  l^{D/2} \psi_a(\br_i)$,
 where  $\br_i$ is the real-space location corresponding to 
 lattice site $i$. The tunneling energy $J$ and $m$ 
 are related through  $J  =  \hbar^2/(2 m l^2)$; 
  $g_0$, $g_{12}$ and the Hubbard interactions through $U  =  g_0/l^D$ and $U_{12}  =  g_{12}/l^D$, respectively. 
 We use $\hbar/J$ as the time unit and choose $U/J = U_{12}/J=0.25$. Using the relation $U/J=\sqrt{32 \pi} a_s/l_0$ from Eq.~(\ref{g0}), this value corresponds to $l_0 \sim 200$nm, which is routinely achieved experimentally.
   We sample the initial state according to the Wigner distribution
 $W(x_\bk, p_\bk)  \sim  \exp (-x_\bk^2/2 \sigma^2_{x, \bk}
  - p_\bk^2/2 \sigma^2_{p, \bk})$, 
 with $\sigma_{x,\bk}^2=1/(2\omega_{\bk} \tanh(\omega_{\bk}/(2T_0)))$
 and  $\sigma_{p,\bk}^2=\omega_{\bk}/(2 \tanh(\omega_{\bk}/(2T_0)))$, 
   where $T_0$ is the initialization temperature, and the Bogoliubov modes 
 $\beta_k  =  \sqrt{\omega_\bk/2} x_\bk + i/\sqrt{2\omega_\bk} p_\bk$.
  From these we construct  $\delta\rho(\br)$ and $\theta(\br)$ 
 and from these  $\psi_1(\br)$, see \cite{densitydistr}.
   $\psi_2(\br_i)$ is initialized with vacuum fluctuations $\langle |\psi_2(\br_i)|^2\rangle=1/2 l^D$. 
   The initial state is propagated by the equations of motion.

In Fig.~\ref{spgf} a)   we show $g_a(x, t) \equiv \langle b_{a, i}^\dagger(t)b_{a, i+x}(t)\rangle$  for $a=1$ for a $250\times 250$ lattice,
 for  $n_0\equiv l^D\rho_0 =5$. We use  $T_0/J = 7$ in the initialization, to let the system equilibrate to a SF with scaling exponent $\tau_i=0.89$.
 We apply the $\pi/2$ pulse at time $J t_{\pi/2} = 1000$; the density drops to half its initial value.  At time $J t_{s}=1200$ we turn off $U_{12}$, see \cite{turnoff} and \cite{contremark}.  
  The evolution after the quench separates into two time domains. The system first relaxes to a metastable SF state, and then changes the functional form of the correlation function from algebraic to exponential on a much longer time scale,  $t \approx 10^3\hbar/J$. 
 To study this quantitatively, we fit the correlation 
function with two functions. We use
 $f_a(x) =  C(\sin(\pi x/N)N/\pi)^{-\tau/4}$
 with $C$ and $\tau$ as fitting parameters to test for algebraic 
 scaling. $N$ is the number of lattice sites in one dimension.
 For large $N$, $f_a(x) \sim |x|^{-\tau/4}$.
  The fitted  exponent $\tau$ is plotted in Fig.~\ref{spgf} b), and for a shorter time evolution and a $100 \times 100$ lattice as the blue dotted line
   in Fig.~\ref{tautime} a), see \cite{suppmat}.
  The constant value of $\tau$ before $t_{\pi/2}$ is $\tau_i$ of our physical initial ensemble.
  For exponential scaling, we use
 $f_e(x)  =  C \exp(-\kappa\sin(\pi x/N)N/\pi)$
 with $C$ and $\kappa$ as fitting parameters.
 We define the fitting errors 
 $E_{a, e}  \equiv  \sum_{i=1}^{N-1} (g(x,t)-f_{a,e})^2$.
 In Fig.~\ref{spgf} c), we show the fitting errors for the data plotted in Fig.~\ref{spgf} a). We recover the time scales discussed above. (i) After the random draw of $x_k$ and $p_k$ and before the $\pi/2$ pulse, the system rapidly equilibrates in few tens of $\hbar/J$ and the two errors fluctuate around a ratio of $E_e/E_a \sim 100$, indicating that $g_1(\bx,t)$ is much better fitted with an algebraic test function. (ii) After the $\pi/2$ pulse and before the quench, the density of species $1$ drops to half its original value after $t_{\pi/2}$ but the gas is still in equilibrium. Both errors  $E_{a,e}$ thus drop by a factor of 4, while maintaining the ratio $\sim 100$. (iii) After the quench, $E_e$ decreases and $E_a$ increases on a long time scale.

 We repeat this calculation for different $\tau_i$, and plot $\tau (t)$ and $E_{a,e}$
 as a function of time and of $\tau_i$, see Fig.~\ref{tautime}.
  For small $\tau_i$,  $\tau (t)$ increases initially and then remains constant below $1$, and  $E_e/E_a \sim 10^2$. For $\tau_i \gtrsim 0.6$, $\tau (t)$ increases above $1$,  and does not remain constant, but shows a slow increase, Fig.~\ref{tautime} a).  $E_a$ grows, and becomes eventually larger
 than $E_e$, as shown in Fig.~\ref{tautime} b) and c), which  indicates the  functional form change 
 of $g_1(\bx, t)$. This demonstrates a dynamic KT transition. 
  We also observe a time regime in which $\tau >1$,
 but the change to exponential scaling has not yet occurred. This
 is the metastable, supercritical SF described in Refs.~\cite{quenchshort, quenchlong}. Further we note that both at initial time and after the second
 quench, the algebraic error briefly spikes up (see also Fig. 2 of Supp. Mat.). This
 is due to the light cone dynamics \cite{quenchshort, quenchlong, cardy} triggered by these two quenches: After each quench, $g_1(x,t)$ is only piece-wise algebraic, i.e. it behaves as $|x|^{-\tau_1/4}$ for $|x|< c(t-t_s)$, and as 
 $|x|^{-\tau_2/4}$ for $|x|> c(t-t_s)$, with differing exponents $\tau_1, \tau_2$;
  thus the fitting error is increased.

\begin{figure*}
\includegraphics[width=17.8cm]{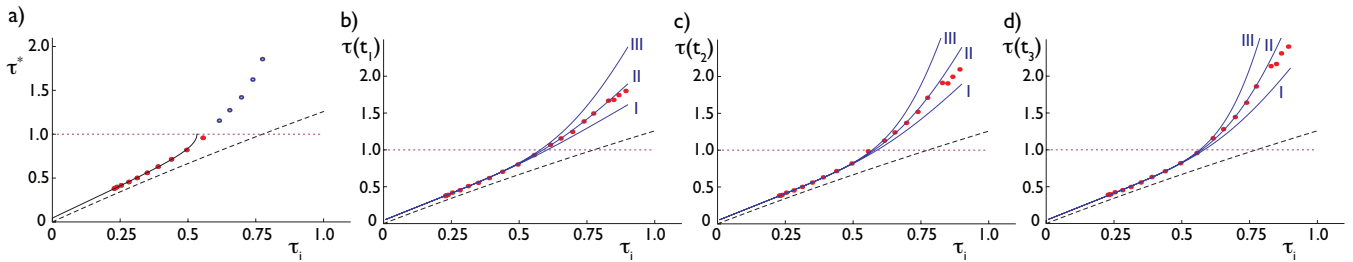}
\caption{\label{fit} a) Numerical data for $\tau$, averaged over $Jt =[2100,3000]$. The red, filled dots have $\tau <1$, the dark circles have $\tau >1$. The black, continuous line is $\tau_1^*$ with the optimal fitting parameters, fitted to the data $\tau <1$. The red, dotted line indicates the
 critical line $\tau =1$. The black, dashed line is $\tau^*_\mathrm{lin}$. 
 b)-d)   Numerical data $\tau$ at times $t_1 = 1500/J$, $t_2 = 2400/J$ and $t_3 = 2970/J$ (b-d)  as a function of $\tau_i$. 
  The blue lines labelled I -- III are 
   in b) $\tau(\ell^*=1.7)$,
 $\tau(\ell^*=2.1)$ and $\tau(\ell^*=2.5)$, respectively.
 In c) $\tau(\ell^*=2.1)$,
 $\tau(\ell^*=2.5)$ and $\tau(\ell^*=2.9)$, respectively, 
  and in d)  $\tau(l^*=2.3)$,
 $\tau(\ell^*=2.7)$ and $\tau(\ell^*=3.1)$, respectively.
  }
\end{figure*}

 We now demonstrate that this transition can be captured with the RG approach derived in Ref.~\cite{quenchlong}. Its key statement is that the transition from the metastable, 'pre-thermalized' state to the fully-thermalized, vortex-deconfined state is described by the flow equations 
\bea
\frac{d y}{d \ell} & = & 2(1-1/\tau)y\label{rg1}; \,\,\, \frac{d \tau}{d \ell}  =  64 \pi^2 \alpha y^2/\tau\label{rg2}
\eea
where  $\alpha$ is a non-universal constant, and $y$ is the vortex fugacity, see  \cite{flow} and \cite{dissipation}. $\ell$ is the flow parameter, related to real time by $t = t_0 e^\ell$, with some time constant $t_0$. We thus have to determine the scaling exponent of the meta-stable state,  referred to as $\tau^*$.  The system has two phononic sectors, corresponding to  symmetric and anti-symmetric superpositions of phase and density excitations. Immediately after the quench, only the symmetric sector
 is thermally activated, whereas the anti-symmetric sector
 only displays vacuum fluctuations. We find that the two sectors thermally equilibrate on a short time scale of $\sim10^2 \hbar/J$,  
  by calculating ratios of 
 $g_s(\bx)\equiv\langle \psi_1(0)^\dagger \psi_{2}(0)^\dagger \psi_{2}(\bx)\psi_1(\bx)\rangle$ and $g_a(\bx)\equiv\langle \psi_1(0)^\dagger \psi_{2}(0) \psi_{2}^\dagger(\bx)\psi_1(\bx)\rangle$, 
such as $g_s(\br)/g(\br)^2$, $g_a(\br)/g(\br)^2$, $g_s(\br) g_a(\br)/g(\br)^4$, etc, which all approach unity. 
 This result differs from the behavior of 1D gases, for which the two sectors stay out-of-equilibrium for a much longer time \cite{kitagawa2011}. 
 To determine $\tau^*$,  we use  energy  conservation.
 For $T$ much larger than the mean-field energy, the total energy scales as $T^2$.  After the quench and equilibration,  the resulting temperature is thus $T/\sqrt{2}$. We use Eq.~(\ref{tau_approx}) for  $\tau$, note that the
 density is reduced by $1/2$, and find $\tau^*_\mathrm{lin}  =  (1/(\sqrt{2} \tau_i) + (1-1/\sqrt{2})*C_0)^{-1}$,
 with $C_0 = (\ln(2 g m/\hbar^2) -1)/4$, for a 2D gas in continuum. However,  $C_0$ will be renormalized for a discretized representation, see Ref.~\cite{NP02}, and we use it as a fitting parameter.  We show $\tau^*_\mathrm{lin}$ in Fig.~\ref{fit} in comparison to the numerical data, with the optimal $C_0$ determined below.
 We use $\tau^*_\mathrm{lin}$ as the initial value for $\tau(0)=\tau^*_\mathrm{lin}$.
  To determine the initial value for the fugacity, we write the flow equation as
 $d (\alpha y^2)/d l  =  4(1-1/\tau)(\alpha y^2)$.
  The quantity $(\tau-1)^2 - 32 \pi^2 \alpha y^2$ 
 is invariant under the flow,   and thus the asymptotic value for $\tau^*$ below the critical point is   $\tau^*(\infty)  =  1 - ((\tau-1)^2 - 32\pi^2\alpha y^2)^{1/2}$. 
 This motivates to use 
 $\tau^*_{1} =  1 - ((\tau^*_\mathrm{lin}-1)^2 - 32\pi^2 A)^{1/2}$ 
  as a fitting function for $C_0$ and $A=\alpha y^2(0)$. 
 In Fig. \ref{fit}  a) we show $\tau^*_1$, with $C_0=0.299$ and $A=2.7\times 10^{-4}$, and the numerical data for $\tau$ averaged over the time range $J t = [2100, 3000]$.   We use these values in our initial conditions.
 We integrate the flow equations to different values $\ell^*$, in 
  particular to $0, 0.1, 0.2, ...,4.0$.
 In Figs.~\ref{fit} (b-d) we show the numerical results for $\tau$
  at the times $t_1 = 1500/J$, $t_2 = 2400/J$ and $t_3 = 2970/J$, each averaged over   a time interval of $60/J$. 
  We determine the value of $\ell^*$ for which $\tau(\ell^*)$
   fits the numerical result the closest. 
    The ratios $t_i/t_j$ are approximately $\exp(\ell_i/\ell_j)$, but we note that because of the logarithmic dependence of $\ell$ on $t$, a large numerical uncertainty is present. 
   We show the optimal $\tau(\ell^*)$, and two
     close-by solutions for visual comparison.
  We find that the RG flow well describes the critical dynamics. 
  
In conclusion, we have presented a realistic experiment to investigate the dynamic KT transition in ultra-cold gases in 2D. We demonstrate that the critical dynamics can be described by the RG approach developed in Ref. \cite{quenchlong}, as it predicts correctly the dynamic evolution found in the TWA simulation. The time evolution of the correlation functions can be detected via interference measurements discussed in Ref. \cite{2Dreview}.
 These predictions and their experimental verification would pave the way to 
  an RG-based theory framework for critical dynamics.

LM acknowledges support from the Landesexzellenzinitiative Hamburg, which is financed by the Science and Research Foundation Hamburg and supported by the Joachim Herz Stiftung. AP was supported by NSF DMR-0907039, AFOSR FA9550-10-1-0110, and the Sloan Foundation.  K-J G and JD were supported by IFRAF and ANR (BOFL project).


\appendix

\section{Supplemental Material}
\begin{figure}[b!]
\includegraphics[width=6.8cm]{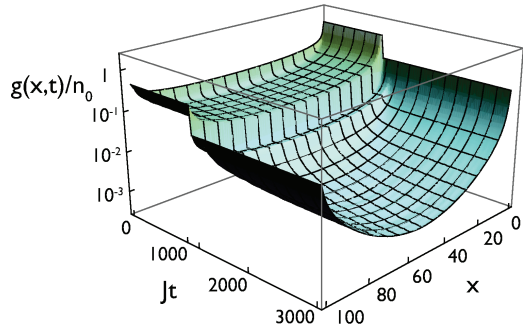}
\caption{\label{suppmatfig1}  
  Single-particle correlation function $g(x,t)$ for $\tau_i=0.89$, for a $100 \times 100$ lattice.}
\end{figure}

We show the single-particle correlation function $g(x,t)$ for a $100\times 100$ lattice, for $\tau_i=0.89$ in Fig.~\ref{suppmatfig1}, for a time interval of
 $Jt = [0, 3000]$.  The $\pi/2$ pulse is applied at $J t_{\pi/2} = 1000$, where the density drops to half, and the quench is applied at $J t_s = 1200$.
As described in the main text, we fit this correlation function with an algebraic and an exponential fitting function. The algebraic fit gives $\tau(t)$ which is depicted
 as  a blue, dashed line in Fig.~3 a) of the main text. 
\begin{figure}[t!]
\includegraphics[width=6.1cm]{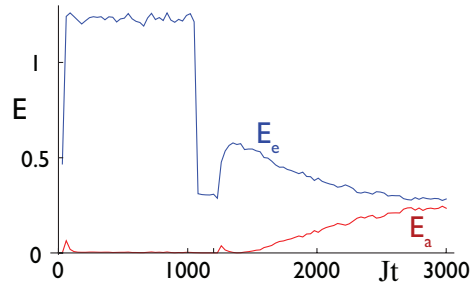}
\caption{\label{suppmatfig2}  
  Algebraic and exponential fitting errors $E_{a,e}$ for $\tau_i=0.89$, for a $100 \times 100$ lattice.}
\end{figure}
 In Fig.~\ref{suppmatfig2} we show the two fitting errors for the data shown in Fig.~\ref{suppmatfig1}. 
 We see the same behavior as for the example in Fig.~2 in the main text, but due to the shorter integration time the intermediate time behavior is better resolved, and because of the smaller system size the numerical uncertainty is reduced. 
 Before the $\pi/2$ pulse, the ratio $E_e/E_a \sim 10^2$, after the pulse each error drops to a quarter of its value while maintaining the same ratio. After the quench at $J t_s = 1200$, the  ratio is initially of that same magnitude, but then changes on a much longer time scale. On an intermediate time scale, there is a metastable regime in which the correlation function is better fitted algebraically, but the exponent $\tau$ is much larger than $1$, i.e. the system is supercritical.

 To construct the data shown in Fig.~3 of the main text, we repeat calculations as shown in Fig.~\ref{suppmatfig1}, for a $100\times 100$ lattice, and for different $\tau_i$, and we determine $\tau(t)$, $E_a(t)$ and $E_e(t)$ from it.


\begin{thebibliography}{99}
\bibitem{densitydistr}
To avoid the rare event of a negative density fluctuation,
 we truncate the distribution of $\delta\rho(\br)$ to
 $-\rho_0$ from below. To ensure that the density expectation value  is $\rho_0$ we truncate
  $\delta\rho(\br)$ to $+\rho_0$ from above.
\bibitem{turnoff}
We use  $U_{12}(t)  =  U (1-\tanh((t-t_s)/\Delta t_s))/2$  with $J \Delta t_s = 5$.  For the example $\Delta t_s = 1.4$ ms.
\bibitem{contremark}
 To describe a 2D  cloud with  $\rho_0=50/(\mu$m$)^2$, we have to choose
 $l=0.32\,\mu$m. For $^{87}$Rb, $\hbar/J$ is then $0.28$ms.
\bibitem{flow}
  Over the temperature range considered here, we ignore the temperature dependence of $y$.  We also ignore the renormalization of the phonon velocity,  see  Ref. \cite{quenchlong}, expected to be only a small correction.
\bibitem{dissipation}
In this coarse-grained formulation, no explicit dissipative term appears, in contrast to Ref. \cite{mitra}. However, a redistribution of energy is included through the conservation of adiabatic invariants, see \cite{quenchlong}.
\end{thebibliography}
\end{document}